\documentclass[12pt,prd,tightenlines,nofootinbib,showpacs,showkeys]{revtex4}
\usepackage{bm}
\usepackage{graphics}
\usepackage{epsfig}
\begin{document}
\title{\begin{flushright}{\rm\normalsize SSU-HEP-07/10\\[5mm]}\end{flushright}
HYPERFINE STRUCTURE OF S-STATES\\ IN MUONIC HELIUM ION}
\author{A. P. Martynenko\footnote{E-mail:~mart@ssu.samara.ru}}
\affiliation{Samara State University, Pavlov str., 1, 443011, Samara, Russia}

\begin{abstract}
Corrections of orders $\alpha^5$ and $\alpha^6$ are calculated in the
hyperfine splittings of $1S$ and $2S$ - energy levels in the ion of
muonic helium. The electron vacuum polarization effects, the nuclear
structure corrections and recoil corrections are taken into account.
The obtained numerical values of the hyperfine splittings
-1334.56 meV ($1S$ state), -166.62 meV ($2S$ state) can be considered
as a reliable estimate for the comparison with the future experimental
data. The hyperfine splitting interval
$\Delta_{12}=(8\Delta E^{hfs}(2S)-\Delta E^{hfs}(1S)) = 1.64$ meV
can be used for the check of quantum electrodynamics.
\end{abstract}


\maketitle

\section{Introduction}

The ion of muonic helium $(\mu ^3_2He)^+$ is the bound state of the
negative muon and helion $(^3_2He$). The lifetime of this simple
atom is determined by the muon decay in a time $\tau_\mu=2.19703 (4)
\times 10^{-6}$ s. An increase of the lepton mass in muonic
hydrogenic atoms as compared with electronic hydrogenic atoms
($m_\mu/m_e=206.7682838(54)$ \cite{MT}) leads to the growth of
three effects in the energy spectrum: the electron vacuum polarization,
the nuclear structure and polarizability, the nuclear recoil effect.
The first effect is important for the spectrum of muonic helium ion 
$(\mu ^3_2He)^+$ because the ratio of the Compton wavelength of the
electron to the Bohr radius of the atom $\mu Z\alpha/m_e\approx 1.45415$
is sufficiently close to the unity. The second effect of the nuclear structure
is of the utmost importance because the muonic wave function has
a significant overlap with the nucleus. Finally, the increase of the
recoil effects is determined by the ratio of the muon and helion masses
$m_\mu/m(^3_2He)\approx 0.0376$ \cite{MT}. Despite the fact that this
ratio is small, it is many times larger than the fine structure constant
$\alpha$ ($\alpha^{-1}=137.03599911(46)$ \cite{MT}). Moreover, some recoil
effects contain the peculiar logarithm of the muon to helion mass ratio
$\ln(m(^3_2He)/m_\mu)\approx 3.28$ what leads to the numerical growth
of the contribution.

High sensitivity of the bound muon characteristics to the distributions
of the nuclear charge and magnetic moment in light muonic atoms
(muonic hydrogen, ions of muonic helium) can be used for more precise
determination of the charge radii of the proton, deuteron, helion and
$\alpha$ - particle \cite{EVSh,EGS,SGK}. Moreover, the measurement of the
hyperfine structure of light muonic atoms allows to obtain more precise
values of the Zemach radii and to improve the accuracy of the theoretical
calculations of the hyperfine splittings in the corresponding electronic
atoms.

Theoretical investigations of the energy levels in light muonic atoms
(the nuclear charge $Z=1,2$) were carried out many years ago in Refs.
\cite{Di,B1,B2,B3,Drake} on the basis of the Dirac equation. In the
calculations of the energy intervals $(2P_{3/2}-2S_{1/2})$, 
$(2P_{1/2}-2S_{1/2})$ in the ions of muonic helium ($^4_2He$, $^3_2He$) 
the different type corrections were considered with the precision
$0.01~meV$. The energy transitions $(2S-2P)$ in the ion of muonic helium
$(\mu ^3_2He)$ were calculated in \cite{B4,B5} with regard to the 
hyperfine structure on the basis of the Dirac equation with the accuracy
$0.1~meV$.

Although muonic atoms $(\mu ^1_1H)$,  $(\mu ^2_1H)$,  $(\mu ^3_2He)^+$,
$(\mu ^4_2He)^+$ could be used for yet another check of quantum 
electrodynamics, the experimental study of the energy levels of these
atoms lags behind the theory. It is appropriate at this point to
recall the experiment for the Lamb shift $(2P-2S)$ measurement in muonic hydrogen
which is carried out many years at PSI (Paul Sherrer Institute) \cite{PSI1,PSI2}
but didn't give while the positive result with the necessary accuracy 30 ppm. 
The only experiment with the successful end was performed on the muon beam at
CERN \cite{CERN,KJ} with muonic helium $(\mu ^4_2He)^+$. Two resonance transitions
with the wavelengths $811.68 (15)$ nm and $897.6(3)$ nm corresponding to the
fine structure intervals $(2P_{3/2}-2S_{1/2})$ and $(2P_{1/2}-2S_{1/2})$
were observed. In a later experiment \cite{Hauser} the resonance transition
in the range $811.4\leq \lambda\leq 812.0$ nm was not observed. So, at present
new measurement of the Lamb shift and the hyperfine structure in the atoms
$(\mu ^3_2He)^+$,  $(\mu ^4_2He)^+$ is required.

There is a need to remark that in the last years the accuracy of the
theoretical investigations of the energy spectra of simple atoms
was increased essentially \cite{EGS}. New QED corrections of order
$\alpha^6$ and $\alpha^7$ in the energy spectra of muonium, positronium,
hydrogen atom and ions of electronic helium were calculated \cite{SGK}. 
For a number of hydrogenic atoms (hydrogen atom, ions of helium)
the comparison of results of the theoretical investigations in QED with the
experiment is difficult because the theoretical error in the calculation
of the nuclear structure and polarizability contributions both for
the Lamb shift and hyperfine structure remains very large and
exceeds considerably the experimental errors. The progress in this
field can be achieved due to new experimental investigations of the
structure and polarizability of the proton and other nucleus
and the use of light muonic atoms.

It is important to keep in mind that all contributions to the energy
spectra of light muonic atoms can be divided into two groups.
The corrections of the first group were obtained in the analytical
form in the study of the energy levels of muonium, positronium and
hydrogen atom. The second group includes numerous corrections of the
electron vacuum polarization, the nuclear structure, recoil effects
which are specific for each muonic atom. The aim of this work
consists in the analytical and numerical calculation of corrections
of orders $\alpha^5$ and $\alpha^6$ in the hyperfine structure of 
$S$-states in the muonic helium ion $(\mu ^3_2He)^+$ on the basis
of the quasipotential method in quantum electrodynamics \cite{M1,M2}.
We consider such effects of the electron vacuum polarization, recoil
and nuclear structure corrections which have the crucial importance
to attain the high accuracy of the calculation. Numerical values
of corrections are obtained with the precision $0.001~meV$. So, the
purpose of our investigation consists in the improvement of the
earlier performed calculations \cite{B1,B2,B3,Drake} of the hyperfine
splitting in the ion of muonic helium and derivation of the reliable 
estimates in the hyperfine structure of $1S$ and $2S$-states which
could be used in conducting a corresponding experiments.
Modern numerical values of fundamental physical constants are
taken from the paper \cite{MT}:
the electron mass $m_e=0.510998918(44)\cdot
10^{-3}~GeV$, the muon mass $m_\mu=0.1056583692(94)~GeV$, the fine
structure constant $\alpha^{-1}=137.03599911(46)$, the helium mass
$m(^3_2He)$ = 2.80839142(24)~GeV, the helium magnetic moment
$\mu_h=-2.127497723(25)$ in the nuclear magnetons, the muon anomalous
magnetic moment $a_\mu=1.16591981(62)\cdot 10^{-3}$.

\section{Effects of one-loop and two-loop vacuum polarization
in the one-photon interaction}

Our approach to the investigation of the hyperfine structure (HFS) in
the muonic helium ion is based on the quasipotential method
in quantum electrodynamics \cite{M3,M4,M5}, where the two-particle
bound state is described by the Schroedinger equation. The basic
contribution to the interaction operator of the muon and helion
for $S$-states is determined by the Breit Hamiltonian \cite{t4}:
\begin{equation}
H_B=H_0+\Delta V_B^{fs}+\Delta V_B^{hfs},~~~H_0=\frac{{\bf p}^2}{2\mu}-\frac{Z\alpha}{r},
\end{equation}
\begin{equation}
\Delta V_B^{fs}=-\frac{{\bf p}^4}{8m_1^3}-\frac{{\bf p}^4}{8m_2^3}+\frac{\pi Z\alpha}{2}
\left(\frac{1}{m_1^2}+\frac{1}{m_2^2}\right)\delta({\bf r})-\frac{Z\alpha}{2m_1m_2r}
\left({\bf p}^2+\frac{{\bf r}({\bf r}{\bf p}{\bf p}}{r^2}\right),
\end{equation}
\begin{equation}
\Delta V_B^{hfs}=\frac{8\pi\alpha\mu_h}{3m_1m_p}
\frac{{\mathstrut\bm\sigma}_1{\mathstrut\bm\sigma}_2}{4}\delta({\bf r}),
\end{equation}
where $m_1$, $m_2$ are the muon and helion masses, $m_p$ is the proton
mass, $\mu_h$ is the helion magnetic moment. The potential of spin-spin
interaction (3) gives the main contribution to the energy of hyperfine
splitting of $S$-states (the Fermi energy). Averaging (3) over the Coulomb
wave functions of $1S$ and $2S$ states
\begin{equation}
\psi_{100}(r)=\frac{W^{3/2}}{\sqrt{\pi}}e^{-Wr},~~~W=\mu Z\alpha,
\end{equation}
\begin{equation}
\psi_{200}(r)=\frac{W^{3/2}}{2\sqrt{2\pi}}e^{-Wr/2}\left(1-\frac{Wr}{2}\right),
\end{equation}
we obtain the following result (the difference of the triplet and singlet states):
\begin{equation}
\Delta E^{hfs}_F(nS)=\frac{8\mu^3 Z^3\alpha^4\mu_h}{3m_1m_pn^3}=
\Biggl\{{{1S:~-1370.725~meV}\atop{2S:~-171.341~meV}},
\end{equation}

The muon anomalous magnetic moment does not enter in Eq.(6). The correction
of the muon anomalous magnetic moment (AMM) to the hyperfine splitting is conveniently
represented by the use the experimental value $a_\mu=1.16591981(62)\cdot 10^{-3}$ \cite{MT}:
\begin{equation}
\Delta E^{hfs}_{a_\mu}(nS)=a_\mu \Delta E^{hfs}_F(nS)=
\Biggl\{{{1S:~-1.598~meV}\atop{2S:~-0.200~meV}}.
\end{equation}

The contribution of the relativistic effects of order $\alpha^6$ 
to the hyperfine structure is also known in the analytical form \cite{EGS}:
\begin{equation}
\Delta E^{hfs}_{rel}(nS)=\left[1+\frac{11n^2+9n-11}{6n^2}(Z\alpha)^2+...\right]
\Delta E^{hfs}_F(nS)=\Biggl\{{{1S:~-0.438~meV}\atop{2S:~-0.078~meV}}
\end{equation}

One-loop electron vacuum polarization correction in the interaction
operator is determined by the following expression in the coordinate
representation \cite{t4}:
\begin{equation}
\Delta V^{hfs}_{1\gamma,VP}(r)=\frac{8\alpha\mu_h}{3m_1m_2}
\frac{{\mathstrut\bm\sigma}_1{\mathstrut\bm\sigma}_2}{4}\frac{\alpha}{3\pi}
\int_1^\infty\rho(s)ds\left[\pi\delta({\bf r})-\frac{m_e^2\xi^2}{r}\right)e^{-2m_e\xi r},
\end{equation}
where $\rho(\xi)=\sqrt{\xi^2-1}(2\xi^2+1)/\xi^4$. To obtain (9) it is necessary
to use the following replacement in the photon propagator:
\begin{equation}
\frac{1}{k^2}\to\frac{\alpha}{3\pi}\int_1^\infty\rho(\xi)d\xi\frac{1}{k^2+4m_e^2\xi^2}.
\end{equation} 
Averaging (9) over the wave functions (4) and (5), we find the correction of order $\alpha^5$
to the hyperfine splitting:
\begin{equation}
\Delta E_{1\gamma,VP}^{hfs}(1S)=\frac{8\mu^3 Z^3\alpha^5\mu_h}{9m_1m_p\pi}\int_1^\infty
\rho(\xi)d\xi\left[1-\frac{4m_e^2\xi^2}{W^2}\int_0^\infty xdxe^{-x\left(1+\frac{m_e\xi}{W}\right)}\right]=
-4.203~meV,
\end{equation}
\begin{equation}
\Delta E_{1\gamma,VP}^{hfs}(2S)=\frac{\mu^3 Z^3\alpha^5\mu_h}{9m_1m_p\pi}\int_1^\infty\rho(\xi)d\xi\times
\end{equation}
\begin{displaymath}
\left[1-\frac{4m_e^2\xi^2}{W^2}\int_0^\infty x(1-\frac{x}{2})^2dxe^{-x\left(1+\frac{2m_e\xi}{W}\right)}\right]=
-0.540~meV.
\end{displaymath}
Changing the electron mass $m_e$ to the muon mass $m_1$ in Eqs. (11), (12),
we obtain the muon vacuum polarization contribution to the hyperfine
splitting of order $\alpha^6$ because the ratio $W/m_1\ll 1$. The corresponding
numerical values are included in Table 1. The contribution of the same order
$\alpha^6$ is given by the two-loop electron vacuum polarization diagrams
in Fig.1(b,c,d). To derive the interaction operator corresponding the
amplitude with two sequential loops in Fig.1(b) we use the double change
(10). In the coordinate representation the interaction operator has the form:
\begin{equation}
\Delta V_{1\gamma,VP-VP}^{hfs}(r)=\frac{8\pi\alpha\mu_h}{3m_1m_p}
\frac{{\mathstrut\bm\sigma}_1{\mathstrut\bm\sigma}_2}{4}\left(\frac{\alpha}
{3\pi}\right)^2\int_1^\infty\rho(\xi)d\xi\int_1^\infty\rho(\eta)d\eta\times
\end{equation}
\begin{displaymath}
\times\left[\delta({\bf r})-\frac{m_e^2}{\pi r(\eta^2-\xi^2)}\left(\eta^4 e^{-2m_e\eta r}-
\xi^4 e^{-2m_e\xi r}\right)\right].
\end{displaymath}
Corresponding correction to the hyperfine splitting of $1S$ and $2S$ levels 
can be presented as a three-dimensional integral over variables
$r$, $\xi$ and $\eta$. After that the integral over $r$ is calculated
analytically and over $\xi$, $\eta$ numerically with the result:
\begin{equation}
\Delta E_{1\gamma,VP-VP}^{hfs}(r)(1S)=\frac{8\alpha^6\mu^3 Z^3\mu_h}{27m_1m_p}
\int_1^\infty\rho(\xi)d\xi\int_1^\infty\rho(\eta)d\eta\times
\end{equation}
\begin{displaymath}
\times\left[1-\frac{4m_e^2}{W^2(\eta^2-\xi^2)}\int_0^\infty e^{-2x}x dx
\left(\eta^4 e^{-2m_e\eta x/W}-\xi^4 e^{-2m_e\xi x/W}\right)\right]=-0.017~meV,
\end{displaymath}
\begin{equation}
\Delta E_{1\gamma,VP-VP}^{hfs}(r)(2S)=\frac{\alpha^6\mu^3 Z^3\mu_h}{27m_1m_p}
\int_1^\infty\rho(\xi)d\xi\int_1^\infty\rho(\eta)d\eta\times
\end{equation}
\begin{displaymath}
\times\left[1-\frac{4m_e^2}{W^2(\eta^2-\xi^2)}\int_0^\infty e^{-x}x dx
\left(\eta^4 e^{-2m_e\eta x/W}-\xi^4 e^{-2m_e\xi x/W}\right)\left(1-\frac{x}{2}\right)^2\right]=-0.002~meV.
\end{displaymath}
In a similar way we calculate the two-loop vacuum polarization contribution
of order $\alpha^6$
shown in Fig.1(c,d). In this case the potential is determined by the relation:
\begin{equation}
\Delta V_{1\gamma,2-loop~VP}^{hfs}(r)=\frac{8\alpha^3\mu_h}{3\pi m_1m_p}
\int_0^1\frac{f(v)dv}{1-v^2}\left[\delta({\bf r})-\frac{m_e^2}{\pi r(1-v^2)}e^{-\frac{2m_er}{\sqrt{1-v^2}}}\right],
\end{equation}
where the function
\begin{equation}
f(v)=v\Bigl\{(3-v^2)(1+v^2)\left[Li_2\left(-\frac{1-v}{1+v}\right)+2Li_2
\left(\frac{1-v}{1+v}\right)+\frac{3}{2}\ln\frac{1+v}{1-v}\ln\frac{1+v}{2}-
\ln\frac{1+v}{1-v}\ln v\right]
\end{equation}
\begin{displaymath}
+\left[\frac{11}{16}(3-v^2)(1+v^2)+\frac{v^4}{4}\right]\ln\frac{1+v}{1-v}+
\left[\frac{3}{2}v(3-v^2)\ln\frac{1-v^2}{4}-2v(3-v^2)\ln v\right]+
\frac{3}{8}v(5-3v^2)\Bigr\},
\end{displaymath}
$Li_2(z)$ is the Euler dilogarithm. Numerical contributions of the operator
(16) to the HFS are included directly in Table 1. The role of the vacuum
polarization effects in the HFS of the muonic helium ion extends further.
There exists a number of contributions where the electron vacuum polarization
enters the potential together with the nuclear structure, recoil and 
relativistic effects in the second order perturbation theory.

\begin{figure}[htbp]
\centering
\includegraphics{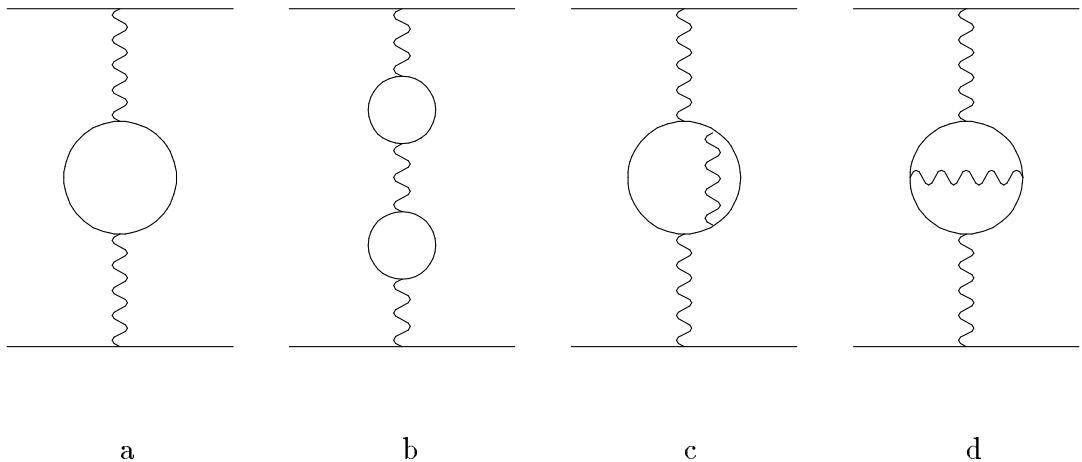}
\caption{Effects of one-loop and two-loop vacuum polarization in the 
one-photon interaction.}
\end{figure}

\section{Effects of one-loop and two-loop vacuum polarization
in the second order perturbation theory}

The second order perturbation theory (PT) corrections in the energy
spectrum of hydrogen-like system are determined by the reduced
Coulomb Green function $\tilde G$ \cite{VP}, whose partial expansion
has the form:
\begin{equation}
\tilde G_n({\bf r}, {\bf r'})=\sum_{l,m}\tilde g_{nl}(r,r')Y_{lm}({\bf n})
Y_{lm}^\ast({\bf n'}).
\end{equation}
The radial function $\tilde g_{nl}(r,r')$ was presented in Ref.\cite{VP} 
in the form of the Sturm expansion in the Laguerre polinomials. The basic
contribution of the electron vacuum polarization to HFS in the second order
PT can be presented as follows (see Fig.2(a)):
\begin{equation}
\Delta E^{hfs}_{SOPT~VP~1}=2<\psi|\Delta V^C_{VP}\cdot \tilde G\cdot\Delta
V_B^{hfs}|\psi>,
\end{equation}
where the modified Coulomb potential
\begin{equation}
\Delta V^C_{VP}(r)=\frac{\alpha}{3\pi}\int_1^\infty\rho(\xi)d\xi\left(-\frac{Z\alpha}{r}\right)
e^{-2m_e\xi r}.
\end{equation}
The potential $\Delta V_B^{hfs}(r)$ is proportional to $\delta({\bf r})$. So,
we need the reduced Coulomb Green function with one zero argument. It was
derived by means of the Hostler representation as a result of the subtraction
of the pole term in Refs.\cite{LM,IK}:
\begin{equation}
\tilde G_{1S}({\bf r},0)=\frac{Z\alpha\mu^2}{4\pi}\frac{e^{-x}}{x}g_{1S}(x),~
g_{1S}(x)=\left[4x(\ln 2x+C)+4x^2-10x-2\right],
\end{equation}
\begin{equation}
\tilde G_{2S}({\bf r},0)=-\frac{Z\alpha\mu^2}{4\pi}\frac{e^{-x/2}}{2x}g_{2S}(x),
\end{equation}
\begin{displaymath}
g_{2S}(x)=\left[4x(x-2)(\ln x+C)+x^3-13x^2+6x+4\right],
\end{displaymath}
where $C=0.5772...$ is the Euler constant, $x=Wr$. Using Eqs.(21), (22)
we can present necessary corrections in the HFS of the ion $(\mu ^3_2He)^+$ 
in the form:
\begin{equation}
\Delta E^{hfs}_{VP~1}(1S)=-\Delta E_F^{hfs}(1S)\frac{2\alpha}{3\pi}(1+a_\mu)\int_1^\infty
\rho(\xi)d\xi
\int_0^\infty e^{-2x\left(1+\frac{m_e\xi}{W}\right)}g_{1S}(x)dx=-9.260~meV,
\end{equation}
\begin{equation}
\Delta E^{hfs}_{VP~1}(2S)=\Delta E_F^{hfs}(2S)\frac{\alpha}{3\pi}(1+a_\mu)\int_1^\infty
\rho(\xi)d\xi\int_0^\infty e^{-x\left(1+\frac{2m_e\xi}{W}\right)}g_{2S}(x)dx
=-0.869~meV.
\end{equation}
The factor $(1+a_\mu)$ is introduced in Eqs.(23), (24) so that these expressions 
contain corrections of orders $\alpha^5$ and $\alpha^6$.

\begin{figure}[htbp]
\centering
\includegraphics{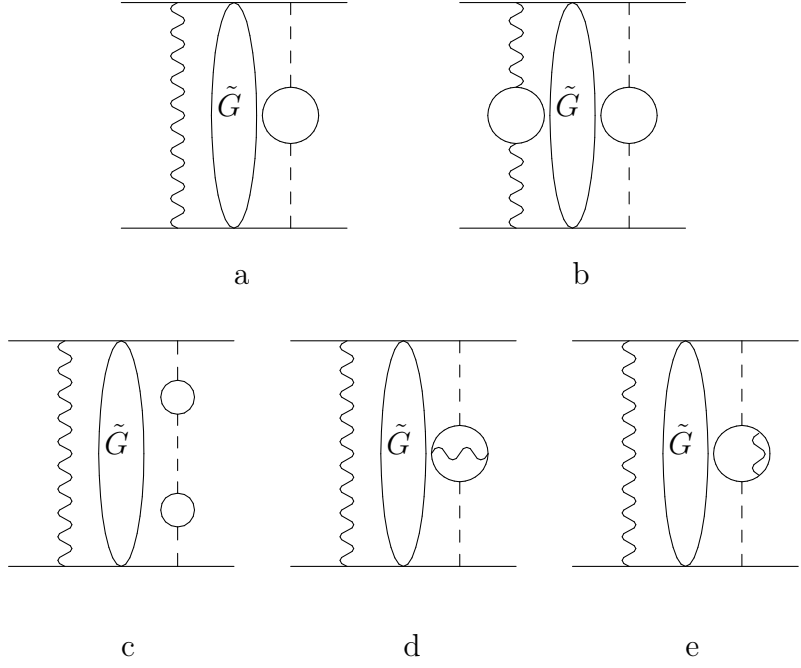}
\caption{Effects of one-loop and two-loop vacuum polarization in the
second order perturbation theory (SOPT). The dashed line represents the Coulomb
photon. $\tilde G$ is the reduced Coulomb Green function.}
\end{figure}

Two-loop contributions in Fig.2(b,c,d,e) are of order $\alpha^6$. Let us consider
first contribution which is determined by the potentials (9), (20), the reduced
Coulomb Green functions (21), (22) and the reduced Coulomb Green functions with both
nonzero arguments. It is convenient to use the compact representation for it
which was obtained in Refs.\cite{LM,KP1}:
\begin{equation}
\tilde G_{1S}(r,r')=-\frac{Z\alpha\mu^2}{\pi}e^{-(x_1+x_2)}g_{1S}(x_1,x_2),
\end{equation}
\begin{displaymath}
g_{1S}(x_1,x_2)=\frac{1}{2x_<}-\ln 2x_>-\ln 2x_<+Ei (2x_<)+\frac{7}{2}-2C-(x_1+x_2)+
\frac{1-e^{2x_<}}{2x_<},
\end{displaymath}
\begin{equation}
\tilde G_{2S}(r,r')=-\frac{Z\alpha\mu^2}{16\pi x_1x_2}e^{-(x_1+x_2)}g_{2S}(x_1,x_2),
\end{equation}
\begin{displaymath}
g_{2S}(x_1,x_2)=8x_<-4x^2_<+8x_>+12x_<x_>-26x^2_<x_>+2x^3_<x_>-4x^2_>-
26x_<x^2_>+23x^2_<x^2_>-
\end{displaymath}
\begin{displaymath}
-x^3_<x^2_>+2x_<x^3_>-x^2_<x^3_>+4e^x(1-x_<)(x_>-2)x_>+4(x_<-2)x_<(x_>-2)x_>
\times
\end{displaymath}
\begin{displaymath}
\times[-2C+Ei(x_<)-\ln(x_<)-\ln(x_>)].
\end{displaymath}
Substituting Eqs.(9), (20), (25) and (26) in  Eq.(19), we obtain two contributions for 
each energy level $1S$ and $2S$:
\begin{equation}
\Delta E^{hfs}_{SOPT~VP~21}(1S)=-\frac{16\alpha^6Z^3\mu^3\mu_h(1+a_\mu)}
{27\pi^2m_1m_p}\int_1^\infty\rho(\xi)d\xi\times
\end{equation}
\begin{displaymath}
\times\int_1^\infty\rho(\eta)d\eta\int_0^\infty
dx e^{-2x\left(1+\frac{m_e\xi}{W}\right)}g_{1S}(x),
\end{displaymath}
\begin{equation}
\Delta E^{hfs}_{SOPT~VP~22}(1S)=-\frac{256\alpha^6Z^3\mu^3\mu_h(1+a_\mu)m_e^2}
{27\pi^2m_1m_pW^2}\int_1^\infty\rho(\xi)d\xi\times
\end{equation}
\begin{displaymath}
\times\int_1^\infty\rho(\eta)\eta^2d\eta\int_0^\infty
x_1dx_1 e^{-2x_1\left(1+\frac{m_e\xi}{W}\right)}\int_0^\infty x_2dx_2
e^{-2x_2\left(1+\frac{m_e\xi}{W}\right)}g_{1S}(x_1,x_2),
\end{displaymath}
\begin{equation}
\Delta E^{hfs}_{SOPT~VP~21}(2S)=\frac{\alpha^6Z^3\mu^3\mu_h(1+a_\mu)}
{27\pi^2m_1m_p}\int_1^\infty\rho(\xi)d\xi\times
\end{equation}
\begin{displaymath}
\times\int_1^\infty\rho(\eta)d\eta\int_0^\infty
\left(1-\frac{x}{2}\right)dx e^{-x\left(1+\frac{2m_e\xi}{W}\right)}g_{2S}(x),
\end{displaymath}
\begin{equation}
\Delta E^{hfs}_{SOPT~VP~22}(2S)=-\frac{2\alpha^6Z^3\mu^3\mu_h(1+a_\mu)m_e^2}
{27\pi^2m_1m_pW^2}\int_1^\infty\rho(\xi)d\xi\times
\end{equation}
\begin{displaymath}
\times\int_1^\infty\rho(\eta)\eta^2d\eta\int_0^\infty
\left(1-\frac{x_1}{2}\right)dx_1 e^{-x_1\left(1+\frac{2m_e\xi}{W}\right)}\int_0^\infty
\left(1-\frac{x_2}{2}\right)dx_2
e^{-x_2\left(1+\frac{2m_e\xi}{W}\right)}g_{2S}(x_1,x_2).
\end{displaymath}
While contributions (27), (28) and (29), (30) are individually divergent,
but their sum is finite. Corresponding numerical values are
represented in Table 1. Corrections from two other diagrams in the
hyperfine structure can be calculated by the relations (23) and (24),
in which the replacement of the potential (20) by the following potentials
should be performed \cite{M5}:
\begin{equation}
\Delta V^C_{VP-VP}(r)=\left(\frac{\alpha}{3\pi}\right)^2\int_1^\infty
\rho(\xi)d\xi\int_1^\infty\rho(\eta)d\eta\left(-\frac{Z\alpha}{r}\right)
\frac{1}{\xi^2-\eta^2}\left(\xi^2 e^{-2m_e\xi r}-\eta^2 e^{-2m_e\eta r}\right),
\end{equation}
\begin{equation}
\Delta V^C_{2-loop~VP}(r)=-\frac{2Z\alpha^3}{3\pi^2 r}\int_0^1\frac{f(v)dv}{(1-v^2)}
e^{-\frac{2m_e r}{\sqrt{1-v^2}}}.
\end{equation}
Omitting further intermediate expressions, which have the general structure
quite similar to (23), (24), we include in Table 1 numerical values of
corrections from potentials (31), (32).
 
\section{Nuclear structure and recoil effects}

The main contribution of the nuclear structure to HFS of the $S$-energy levels
including the Zemach correction, is determined by two-photon exchange diagrams
(see Fig.3). We consider that the nuclear charge and magnetic moment are
distributed in the space with definite densities. The vertex operator of the
nucleus $^3_2He$ contains the electric $G_E$ and magnetic $G_M$ form factors
which determine the interaction of the nucleus with the electromagnetic field. 
To calculate the nuclear structure correction of order $\alpha^5$ we use 
the equation from Ref.\cite{M4} (the abbreviation "str" is used to designate the 
nuclear structure correction):
\begin{equation}
\Delta E_{str}^{hfs}=-\frac{(Z\alpha)^5}{3\pi m_1m_2n^3}\delta_{l0}\int_0^\infty\frac{dk}{k}V(k),
\end{equation}
\begin{displaymath}
V(k)=\frac{2F_2^2k^2}{m_1m_2}+\frac{\mu}{(m_1-m_2)k(k+\sqrt{4m_1^2+k^2})}
\Biggl[-128F_1^2m_1^2-128F_1F_2m_1^2+16F_1^2k^2+
\end{displaymath}
\begin{displaymath}
+64F_1F_2k^2+16F_2^2k^2+\frac{32F_2^2m_1^2k^2}{m_2^2}+\frac{4F_2^2k^4}
{m_1^2}-\frac{4F_2^2k^4}{m_2^2}\Biggr]+\frac{\mu}{(m_1-m_2)k(k+
\sqrt{4m_2^2+k^2})}\times
\end{displaymath}
\begin{displaymath}
\times\left[128F_1^2m_2^2+128F_1F_2m_2^2-16F_1^2k^2-64F_1F_2k^2-48F_2^2k^2\right].
\end{displaymath}
To remove infrared divergence in Eq.(33) we must take into account
the contribution of the iterative term of the quasipotential
to the HFS of the atom $(\mu ^3_2He)^+$:
\begin{equation}
\Delta E_{iter,str}^{hfs}=-<V_{1\gamma}\times G^f\times
V_{1\gamma}>_{str}^{hfs}=-\frac{64}{3}\frac{\mu^4Z^4\alpha^5
\mu_h}{m_1m_p\pi n^3}\int_0^\infty\frac{dk}{k^2},
\end{equation}
where the angle brackets denote the averaging of the interaction
operator over the Coulomb wave functions and the index "hfs"
is indicative of the hyperfine part in the iterative term of the
quasipotential. $V_{1\gamma}$ is the quasipotential of the one-photon
interaction, $G^f$ is the free two-particle propagator. The integration
in Eqs.(33) and (34) can be done by means of
the dipole parameterization for the Pauli form factor
$F_1$ and the Dirac form factor $F_2$ \cite{Friar,Sick}.
The parameter for this parameterization $\Lambda^2$ can be related
with the nuclear charge radius $r_N$: $\Lambda^2=12/r_N^2$. Numerical
value of $r_N=1.844\pm 0.045$ fm is taken from Ref.\cite{B3}. 
Since the dependence on the principal quantum number $n$ in Eq.(33) 
is determined by the factor $1/n^3$, the numerical values of the
nuclear structure corrections for the states $1S$ and $2S$
\begin{equation}
\Delta E_{str}^{hfs}=\Biggl\{{{1S:~48.376~meV}\atop{2S:~6.047~meV}}
\end{equation}
are cancelled in the special hyperfine splitting interval 
$\Delta_{12}=(8\Delta E^{hfs}_{str}(2S)- \Delta E^{hfs}_{str}(1S))$. 
So, the calculation of the interval $\Delta_{12}$ is free of the
uncertainty connected with the nuclear structure at least in the
leading order. The value of the correction (33) is dependent on the
form of distributions $G_{E,M}$. The replacement of the dipole
parameterization by the Gaussian model leads to the change
of the numerical value (35) by 2 per cent.

\begin{figure}[htbp]
\centering
\includegraphics{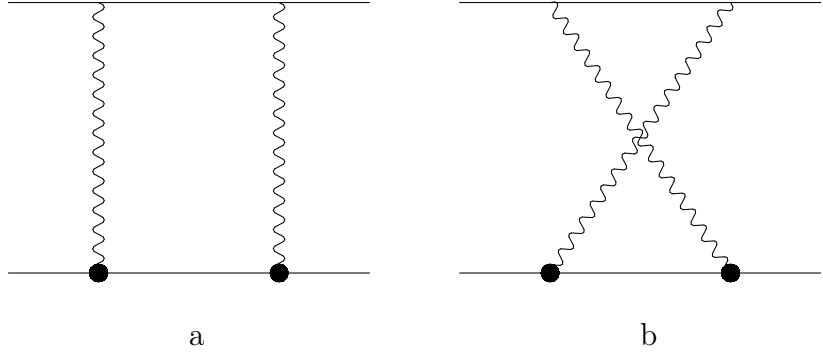}
\caption{Effects of the nuclear structure of order $\alpha^5$.}
\end{figure}

The sixth order over $\alpha$ contribution to the HFS shown in Fig.4
contains both the nuclear structure and the vacuum polarization effects.
Using the substitution (10) and Eq.(33) it can be written as follows:
\begin{equation}
\Delta E_{str,VP}^{hfs}=-\frac{2\alpha(Z\alpha)^5\mu^3}{m_1m_2\pi^2 n^3}
\int_0^\infty V_{VP}(k)dk\int_0^1\frac{v^2\left(1-\frac{v^2}{3}\right)dv}
{k^2(1-v^2)+4m_e^2},
\end{equation}
where the potential $V_{VP}(k)$ differs on the expression $V(k)$ in Eq.(33)
by the additional factor $k^2$. The amplitude contribution (36) to the energy
spectrum should be augmented by two iterative terms which are presented
in Fig.4(c,d):
\begin{equation}
\Delta E_{iter,str~VP}^{hfs}=-2<V^C\times G^f\times\Delta V_{VP}^{hfs}>^{hfs}=
-2<V^C_{VP}\times G^f\times\Delta V_B^{hfs}>^{hfs}=
\end{equation}
\begin{displaymath}
=-\Delta E_F^{hfs}\frac{4\mu\alpha(Z\alpha)}{m_e\pi^2}\int_0^\infty dk\int_0^1\frac{v^2\left(
1-\frac{v^2}{3}\right)dv}{k^2(1-v^2)+1}.
\end{displaymath}
Numerically the sum of corrections (36) and (37) is equal
\begin{equation}
\Delta E^{hfs}_{str,VP}+2\Delta E_{iter,str~VP}^{hfs}=
\Biggl\{{{1S:~0.760~meV}\atop{2S:~0.095~meV}}
\end{equation}

\begin{figure}[htbp]
\centering
\includegraphics{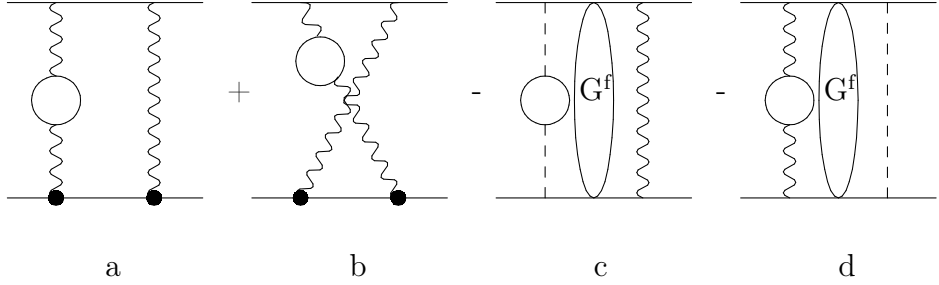}
\caption{Nuclear structure and vacuum polarization effects of order
$\alpha^6$. The dashed line represents the Coulomb photon.
$G^f$ is the free two-particle propagator.}
\end{figure}

\begin{table}
\caption{\label{t1}Hyperfine structure of $1S$ and $2S$ states in the ion
of muonic helium $(\mu~^3_2He)^+$. $\Delta_{12}= (8\Delta
E^{hfs}(2S)-\Delta E^{hfs}(1S))$.}
\bigskip
\begin{ruledtabular}
\begin{tabular}{|c|c|c|c|c|}  \hline
Contribution to HFS &$1S$,~meV& $2S$,~meV & Interval $\Delta_{12}$& Reference   \\   \hline
The Fermi energy & -1370.725 & -171.341 & 0 & (6), \cite{EGS,SGK}  \\  \hline
Muon AMM correction of orders $\alpha^5$ and $\alpha^6$ & -1.598 & -0.200 & 0 & (7), \cite{EGS,SGK} \\ \hline
Relativistic correction of order $\alpha^6$  & -0.438  & -0.078 & -0.183  & (8), \cite{EGS} \\ \hline
One-loop VP contribution in $1\gamma$ &  -4.203   &  -0.540    &-0.119      &  (11)-(12)     \\
interaction of order $\alpha^5$                   &     &      &     &     \\   \hline
Two-loop VP contribution in $1\gamma$ &  -0.050   &  -0.004    &   0.016   &  (14)-(16)     \\
interaction of order $\alpha^6$                   &     &      &     &     \\   \hline
One-loop muon VP contribution &  -0.052   &  -0.007    &0      &  (11)-(12)     \\
in $1\gamma$ interaction of order $\alpha^6$                   &     &      &     &     \\   \hline
One-loop VP contribution in the &  -9.260   &  -0.869    &2.305      &  (23)-(24)     \\
second order PT of order $\alpha^5$                   &     &      &     &     \\   \hline
Two-loop VP contribution in the &  -0.105   &  -0.010    &0.022      &  (27)-(30)     \\
second order PT of order $\alpha^6$                   &     &      &     &     \\   \hline
Nuclear structure correction of order $\alpha^5$  & 48.376  & 6.047 & 0  & (33),(34) \\ \hline
Contribution of VP and nuclear &  0.760   &  0.095    &0      &  (38)     \\
structure of order $\alpha^6$ &     &      &     &     \\   \hline
Nuclear structure correction &  2.553   &  0.272   &-0.377     &  (41)-(42),(45)     \\
of order $\alpha^6$ &     &      &     &     \\   \hline
Nuclear structure and muon  ¨&  -0.145   &  -0.018   &0      &  (46), \cite{SGK1}     \\
self-energy correction of order $\alpha^6$ &     &      &     &     \\   \hline
Nuclear recoil correction &  0.330   &  0.038   &-0.026      &  (47)-(48),\cite{S1}     \\
of order $\alpha^6$ &     &      &     &     \\   \hline
Summary contribution & -1334.560  &-166.615  &  1.638   &    \\  \hline
\end{tabular}
\end{ruledtabular}
\end{table}
 
There exists another contribution of the nuclear structure in the
second order PT, which is determined by the hyperfine term of the
Breit Hamiltonian and the operator of the one-photon interaction
(see Fig.5(b))
\begin{equation}
\Delta V_{str}=\frac{2\pi(Z\alpha)}{3}r_N^2\delta({\bf r}).
\end{equation}
The nuclear structure effects are taken into account in Eq.(39)
in terms of the nuclear charge radius $r_N$. This contribution
has the form:
\begin{equation}
\Delta E^{hfs}_{str~SOPT}(nS)=2<\psi_n|\Delta V_B^{hfs}\cdot\tilde G
\cdot \Delta V_{str}|\psi_n>=\frac{4\pi(Z\alpha)}{3}\Delta E_F^{hfs}(nS)r_N^2\tilde G(0,0).
\end{equation}
The value of the reduced Coulomb Green function at zero arguments in
the coordinate representation $\tilde G(0,0)$ is divergent. The reason of the
appeared divergence lies in the expansion of the potentials in Eq.(40)
at small relative momenta and further integration (40) at all values
of relative momenta. To calculate the quantity $\tilde G(0,0)$ the dimensional
regularization can be useful \cite{Hoang,CMY,SGK1}. Subtracting the iterative
term $2<\psi_n|\Delta V_B^{hfs}\cdot G^f\Delta V_{str}|\psi_n>$ from Eq.(40)
we obtain the following result:
\begin{equation}
\Delta E^{hfs}_{str~SOPT}(1S)=\frac{4}{3}(Z\alpha)^2m_1^2r_N^2\Delta E_F^{hfs}(1S)
\left[\ln(Z\alpha)-\frac{3}{2}\right],
\end{equation}
\begin{equation}
\Delta E^{hfs}_{str~SOPT}(2S)=\frac{4}{3}(Z\alpha)^2m_1^2r_N^2\Delta E_F^{hfs}(2S)
\left[\ln(Z\alpha)-\ln 2\right].
\end{equation}
One further contribution of the sixth order over $\alpha$ can be derived from
the one-photon amplitude (Fig.5(a)) expanding the nuclear magnetic formfactor
at small values of the relative momenta. As a result the potential of
the hyperfine interaction (3) in the coordinate representation gains the
additional term
\begin{equation}
\Delta V_{1\gamma~str}^{hfs}(r)=-\frac{4\pi\alpha(1+a_\mu)}{9m_1m_p}
r_M^2\frac{{\mathstrut\bm\sigma}_1{\mathstrut\bm\sigma}_2}{4}
\nabla^2\delta({\bf r}),
\end{equation}
where $r_M$ is the nuclear magnetic radius. For the averaging the
operator (43) over the Coulomb wave functions we use the following
relation
\begin{equation}
\int \nabla^2\delta({\bf r})d{\bf r}|\psi_n({\bf r})|^2=2\left(\psi(0)
\nabla^2\psi(0)+\left(\frac{d\psi_n}{dr}\right)^2|_{r=0}\right),
\end{equation}
and the value $\nabla^2\psi(0)=\psi(0)\mu^2(Z\alpha)^2\frac{3+2(n^2-1)}{n^2}$
\cite{M1,KM1}. As a result we obtain the important correction proportional to
the nuclear magnetic radius:
\begin{equation}
\Delta E_{1\gamma,str}^{hfs}(nS)=-\frac{4}{3}(Z\alpha)^2\mu^2r_M^2\Delta E_F^{hfs}(nS)
\frac{1-n^2}{4n^2}.
\end{equation}
We have included in Table 1 the total nuclear structure contribution which
is determined by expressions (41), (42) and (45) for $1S$ and $2S$ energy
levels at $r_M\approx r_N$. Let us write here also the corection of order
$\alpha^6$ connected with the nuclear structure and the muon self energy 
\cite{SGK1}:
\begin{equation}
\Delta E_{str~SE}^{hfs}=\frac{5}{2}\frac{\alpha(Z\alpha)}{\pi}m_1R_Z\Delta E_F^{hfs}=
\Biggl\{{{1S:~-0.145~meV}\atop{2S:~-0.018~meV}},
\end{equation}
where $R_Z$ is the Zemach radius.

\begin{figure}[htbp]
\centering
\includegraphics{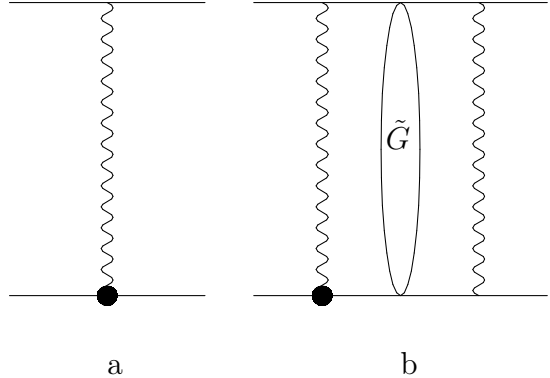}
\caption{Nuclear structure effects of order $\alpha^6$ in the one-photon
interaction and the second order PT. $\tilde G$ is the reduced Coulomb
Green function.}
\end{figure}

One part of the recoil corrections to the HFS is accounted in the calculation
of the diagrams in Figs.3-4. Thus the leading order recoil contribution
$(Z\alpha)(m_1/m_2)\ln(m_1/m_2)\Delta E_F^{hfs}$ is contained in the potential  (34). 
The recoil correction of order $(Z\alpha)^2(m_1/m_2)\Delta E_F^{hfs}$ for the ground state
HFS in the hydrogen atom was calculated in Refs.\cite{BY,SGK1} and for the
hyperfine interval $\Delta_{12}$ in Ref.\cite{S1}. Using these results we can
present the analytical expressions for the recoil corrections and their
numerical values for the HFS of $1S$ and $2S$ -states in the form:
\begin{equation}
\Delta E_{rec}^{hfs}(1S)=(Z\alpha)^2\frac{\mu^2}{m_1m_2}\Delta E_F^{hfs}(1S)\Bigl[
-\frac{17}{12}+\frac{25}{3\zeta}+\frac{31\zeta}{72}+
\end{equation}
\begin{displaymath}
+\ln 2\left(\frac{1}{2}-
\frac{23}{2\zeta}-\frac{11\zeta}{8}\right)+\ln\left(\frac{1}{Z\alpha}\right)
\left(-\frac{3}{2}+\frac{7}{2\zeta}+\frac{7\zeta}{8}\right)\Bigr]=0.330~meV,
\end{displaymath}
\begin{equation}
\Delta E_{rec}^{hfs}(2S)=(Z\alpha)^2\frac{\mu^2}{m_1m_2}\Delta E_F^{hfs}(2S)\Bigl[
-\frac{265}{96}+\frac{821}{96\zeta}-\frac{809\zeta}{1152}+
\end{equation}
\begin{displaymath}
+\ln 2\left(1-\frac{12}{\zeta}-\frac{\zeta}{2}\right)+\ln\left(\frac{1}{Z\alpha}\right)
\left(-\frac{3}{2}+\frac{7}{2\zeta}+\frac{7\zeta}{8}\right)\Bigr]=0.038~meV,
\end{displaymath}
where $\zeta=2m_2\mu_h/m_pZ$.

\section{Summary and Conclusion}

In this work various QED corrections, effects of the nuclear structure and recoil
of orders $\alpha^5$ and $\alpha^6$ have been calculated for the hyperfine splittings of 
$1S$ and $2S$ energy levels in the ion of muonic helium $(\mu ^3_2He)^+$. The investigation
of the energy structure of $1S$ and $2S$ states in this atom has the clear 
experimental prospect. Contrary to earlier performed studies of the energy spectra
of light muonic atoms in Refs.\cite{B1,B2,B3} we use the three-dimensional quasipotential
method for the description of the muon and helion bound state. All corrections considered
here can be separated into two groups. The first group consists of the contributions
which are specific for muonic helium ion. Primarily they are connected with the
effects of the electron vacuum polarization. In our study these contributions are
presented in the integral form and obtained numerically. The corrections known in 
the analytical form from the calculations of the hyperfine structure of muonium
and hydrogen atom enter in the second group \cite{EGS}. Numerical values of all
corrections are presented in Table 1. It contains also several basic references
on the papers where the precision calculations of the HFS of simple atoms were
considered. Other references can be founded in the review articles \cite{EGS,SGK}.

As mentioned above, the hyperfine structure of light exotic atoms was
investigated on the basis of the Dirac equation many years ago in 
Refs.\cite{B4,B5}. The energies of the transitions $(2S-2P)$ were obtained for
the muonic hydrogen and ion of muonic helium $(\mu ^3_2He)^+$. In this calculation
only basic contributions to the HFS with the precision 0.1 meV
were accounted. It follows from the Table 2 of Ref.\cite{B5}, that
the energies of the transitions $(^1S_{1/2}-{^3P_{1/2}})$ and $(^3S_{1/2}-{^3P_{1/2}})$
are equal correspondingly 1167.3 meV and 1334.1 meV resulting the definite
value of the hyperfine splitting of $2S$-state: -166.8 meV. The total value
of the HFS of $2S$ energy level entering in our Table 1 -166.615 meV
is in good agreement with this result. So, the calculation of the HFS
in the ion of muonic helium performed in this work improves the obtained
earlier result in \cite{B5} for the $2S$ state by the calculation of order
$\alpha^6$ corrections and gives new result for the hyperfine splitting of
$1S$ state. The estimate of the next to considered order contribution over
$\alpha$ has the form: $\alpha^3\ln(1/\alpha)\Delta E_F^{hfs}(1S)\approx 0.003$ meV. 
Despite the fact that all contributions in Table 1 are written with the
accuracy 0.001 meV, the precision of our calculation of the HFS is not so high.
The reason is that the nuclear structure correction of order $\alpha^5$ 
has the largest theoretical uncertainty associated with the errors in the
measurement of electromagnetic form factors for the nucleus $^3_2He$. When the
dipole parameterization for the form factors is used the value of the
theoretical uncertainty is determined by the error of the nuclear charge radius:
$r_N(^3_2He)=1.844\pm 0.045$ fm. As a result the theoretical error can reach
near $\pm 1.5$ meV for $1S$-level and $\pm 0.20$ meV for $2S$-level.
Nuclear corrections to the HFS connected with the motion of the nucleons
forming the nucleus $^2H$, $^3H$, $^3He$ of light hydrogen-like atoms
were studied in Refs.\cite{Friar2,Low}. Another source of the uncertainty
is connected with the nuclear polarizability effect 
\cite{IBK1,M6,M7,M8,IBK2,CEC1,CEC2,CEC3,KP2007}.
It demands further investigation on the basis of the experimental data
on the polarized lepton scattering by the nucleus $^3_2He$. The value
of the nuclear polarizability contribution to the HFS of the muonic
helium ion can amount to several meV. The nuclear polarizability contribution 
should be considered in the combination with the nuclear corrections because
both effects are associated with the interaction of the multinucleon system
with electromagnetic field. The hyperfine splitting interval $\Delta_{12}$
has not uncertainties conditioned by the nuclear structure and polarizability
corrections. So, the obtained value of the interval $\Delta_{12}=1.638$ meV can be used
for the check of QED predictions in the case of muonic helium ion with the
accuracy 0.01 meV.

\begin{acknowledgments}
The author is grateful to R.N.Faustov for fruitful discussions. The final
part of the work was carried out in the Humboldt University in Berlin and
I am grateful colleagues from Institute of Physics for warm hospitality. 
This work was supported in part by the Russian Foundation for Basic 
Research (Grant No.06-02-16821).
\end{acknowledgments}

\end{document}